\documentclass[singlecolumn,showpacs,superscriptaddress,notitlepage]{revtex4-1}
\usepackage{amsmath,amssymb,amsfonts,graphics,graphicx,dcolumn,bm}
\usepackage[toc,page]{appendix}
\usepackage{natbib}
\usepackage{geometry}
\usepackage{qtree}
\usepackage{chngcntr}

\usepackage[ddmmyyyy,hhmmss]{datetime}

\newcommand{\aalto}
{\affiliation{Department of Computer Science, Aalto University School of Science, 00076, Finland}}

\newcommand{\oxford}
{\affiliation{Department of Experimental Psychology, University of Oxford, OX1 3UD, United Kingdom}}

\begin{document}

\title{Quantifying gender preferences across humans lifespan}
\author{Asim Ghosh}
\email[Email:]{asim.ghosh@aalto.fi}
\aalto
\author{Daniel Monsivais}
\email[Email:]{daniel.monsivais-velazquez@aalto.fi}
\aalto
\author{Kunal Bhattacharya}
\email[Email:]{kunal.bhattacharya@aalto.fi}
\aalto
\author{Robin I. M. Dunbar}
\email[Email: ]{robin.dunbar@psy.ox.ac.uk}
\oxford
\aalto
\author{Kimmo Kaski}
\email[Email:]{kimmo.kaski@aalto.fi}
\aalto
\oxford

\begin{abstract}
In human relations individuals' gender and age play a key role in the structures and dynamics of their social arrangements. In order to analyze the gender preferences of individuals in interaction with others at different stages of their lives we study a large mobile phone dataset. To do this we consider four fundamental gender-related caller and callee combinations of human interactions, namely male to male, male to female, female to male, and female to female, which together with age, kinship, and different levels of friendship give rise to a wide scope of human sociality. Here we analyse the relative strength of these four types of interaction using a large dataset of mobile phone communication records. Our analysis suggests strong age dependence for an ego of one gender choosing to call an individual of either gender. We observe a strong opposite sex bonding across most of their reproductive age. However, older women show a strong tendency to connect to another female that is one generation younger in a way that is suggestive of the \emph{grandmothering effect}. We also find that the relative strength among the four possible interactions depends on phone call duration. For calls of medium and long duration, opposite gender interactions are significantly more probable than same gender interactions during the reproductive years, suggesting potential emotional exchange between spouses. By measuring the fraction of calls to other generations we find that mothers tend to make calls more to their daughters than to their sons, whereas fathers make calls more to their sons than to their daughters. For younger people, most of their calls go to same generation alters, while older people call the younger people more frequently, which supports the suggestion that \emph{affection flows downward}.
\end{abstract}

\maketitle

\section{Introduction}
In social interactions between humans, gender and age play a key role in the communities and social structures they form and the dynamics therein. For the caller-callee interactions in mobile communication there are four fundamental possibilities, namely male to male, male to female, female to male, and female to female, which together with age, kinship, and different levels of friendships affect the strengths of social interactions, giving rise to a wide scope of human sociality. The studies of primate brain size and its relation to their average social group size suggest that humans are able to maintain of the order of 150 stable relationships (Dunbar number) \cite{dunbar1992neocortex,de2011dunbar,gonccalves2011modeling,maccarron2016calling}. In addition the Social Brain hypothesis suggests that on the basis of emotional closeness human social networks can be divided into four cumulative layers of 5, 15, 50 and 150 individuals, respectively \cite{hill2003social}. As the concept of emotional closeness is hard to quantify, we follow previous studies and assume that it can be equated to the frequency of communication between an ego and an alter. This makes the concept quantifiable such that one can observe how much an ego shares social resources with alters of different gender and age. 

Over the past decade or so, much research on human communication patterns has been done by using ``digital footprints'' data from modern communication technologies such as mobile phone calls and text messages as well as social media like Facebook, and Twitter \cite{blondel2015survey,ellison2007benefits,kwak2010twitter}. Of these the mobile phone communication data of call detail records (CDR's) has turned out to help us in getting insight into the structure and dynamics of social network, human mobility and behavioral patterns in much finer details than before \cite{blondel2015survey}. It has also revealed how microscopic properties related to individuals translate to macroscopic features of their social organisation such as networks. As a result of these studies we now have quite a good understanding of a number of structural properties of human social networks such as degree, strength, clustering coefficient, community structure, and motifs
\cite{onnela2007structure,onnela2007analysis,kovanen2011temporal}. 

Apart from these basic structural properties of networks, more recent studies have given us insight into a number of other aspects of social networks, namely their dependence on temporal, geographic, demographic, and behavioral factors of individuals in the network. One such observation pertains to the shifting patterns of human communication across the reproductive period of their lives, which appears to reflect parental care \cite{palchykov2012sex,david2015communication}. Another is a study using the postal code information in the data to show that the tie strength is related to  geographical distance \cite{onnela2011geographic}. In addition, it has been shown that there is a universal pattern of time allocation to differently ranked alters \cite{saramaki2014persistence}. Finally, recent studies indicate variation in connections and the number of friends with the age and gender \cite{bhattacharya2015sex,dong2014inferring}.

In the present study, we focus on measuring the relative strengths of the four possible pairwise ego-alter interactions over their lifespans as a function of ego's age. From the point of view of call initiation, we find that females play a more active role during their reproductive years as well as during their grandmothering period. We also find that females nearing the age of marriage are strongly linked to their best male friends. Similarly, females of grandmothering age are found to give more attention to their children, while males up to the age of 50 years still keep stronger connection with their spouses of slightly younger age. Furthermore, the fraction of calls to individuals of different generations indicates that mothers call their daughters more than their sons, whereas fathers call their sons more than their daughters.  For younger people, most of their calls go to same generation alters, whereas older people call younger people more frequently, implying that affection flows downward.

\section{Methodology}
In this study we analyse mobile phone communication records of a particular European mobile service provider containing time series of call detail records or CDR's of caller-callee pairs. This dataset also includes demographic information of age, gender, and common locations etc. of the callers. By using the gender information we measure the relative strengths for the four basic calling pattern such that we count the total fraction of calls for the caller-callee pairs of the same or of different genders by assuming a cutoff for the minimum call duration. We analyse all the CDR's for the year 2007 on a month-by-month basis for more than 2.4 million subscribers where both the caller's and the callee's demographics are known, totaling over 30 million calls. Since datasets of this kind are susceptible to error due to multiple subscriptions, we filtered out customers who have multiple subscriptions under the same contract number.

\begin{figure}[t]
\includegraphics[width=16cm]{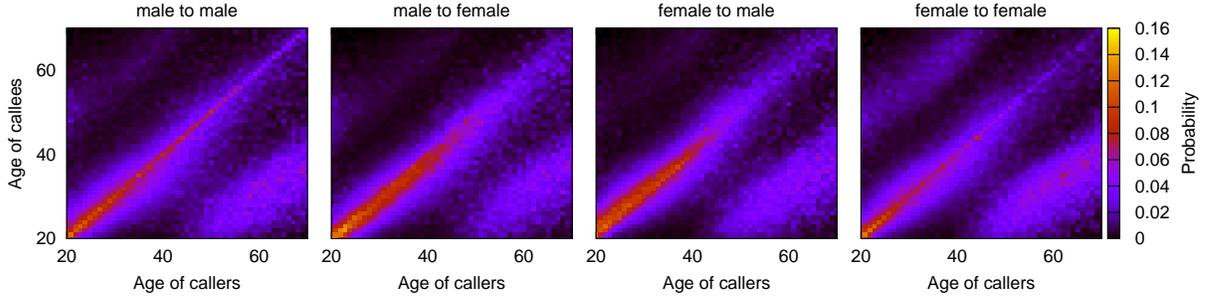}
\caption{Snapshots of all four possible ways of interaction (male to male, male to female, female to male and female to female) are shown by considering only calls of duration more than one minute and caller and callee ages from 20 years to 70 years. For given age and gender of callers, the calling probability is calculated as the ratio between total number of calls to callees of specific age and the total number of calls to all callees of ages between 20 and 70 years.} 
\label{fig1}
\end{figure}

\section{Results}
In order to analyze the gender preferences of individuals in interaction with others at different stages of their lives we choose the age of the caller and count the total number of calls within a time window. We apply a threshold for minimum call duration, such that calls shorter than the threshold value are considered not to be indicative of emotional closeness while calls longer than that are taken to indicate a meaningful emotional or social exchange relationship between the caller and the callee. Then we calculate the relative probabilities for the four possible types of caller to callee interaction. As it is difficult to decide a priori where the borderline between meaningless and meaningful is, we will vary the threshold for minimum call duration in measuring how the probabilities of the four ways of interaction vary with age and gender of the callers. 

In Fig. \ref{fig1} we show snapshots of the four interaction types for a call duration threshold of one minute. By considering callers and callees between 20 and 70 years of age we have calculated the calling or interaction probabilities for the same and different gender pairs. Here the probability is determined as the ratio between the total number of calls to the specific age callee and the total number of calls to all the callees within the age range of 20 to 70 years. From these communication patterns, the signature of a generation gap becomes evident.  For example, the calling pattern between two females exhibits a rather clear signature of being triple lobed with the side lobes separated from the same age group center lobe by a generation gap up and down. This is indicative of frequent interactions between mothers and their daughters over the two generation gaps. 

\begin{figure}[t]
 \includegraphics[width=12cm]{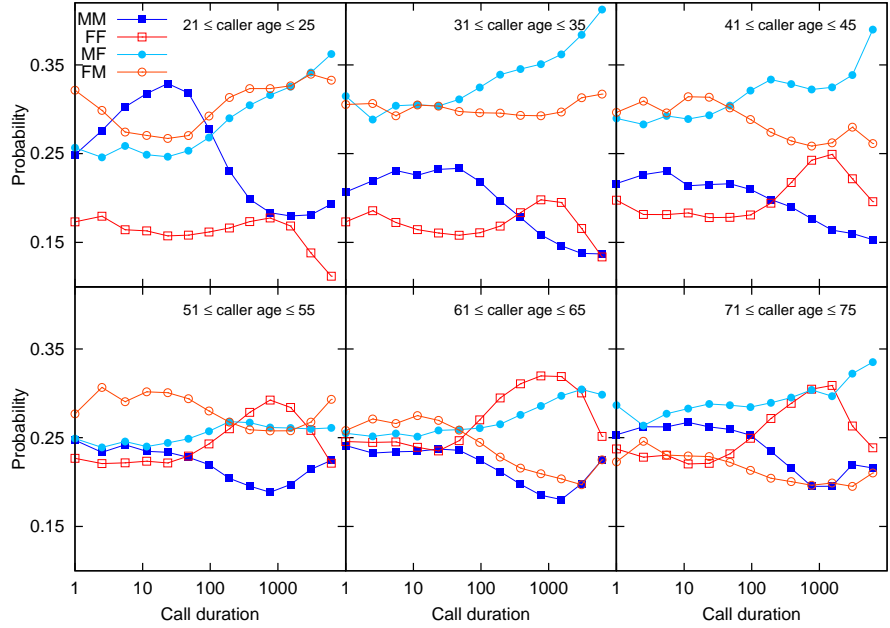}
 \caption{Relative probabilities of the four possible ways of interaction between same or different gender caller - callee pairs as a function of the call duration for six different age groups of callers, presented as panels of 21-25, 31-35, 41-45, 51-55, 61-65, and 71-75 years old. The relative ranking of these four possibilities is dependent on the age of the caller and the duration of calls.} 
\label{fig2}
\end{figure}

In Fig. \ref{fig2} we show the probabilities of the four possible ways of interaction for caller-callee pairs as a function of the call duration for different age groups of callers, i.e. 21-25, 31-35, 41-45, 51-55, 61-65, and 71-75 years. Here we find that the relative ranking between them is strongly dependent on  call duration. At younger ages (21-25 years), male to male (MM) calls tend to be relatively short, with  interactions  peaking  around  10  secs  and  being  of  the  highest  rank  up to 100 secs then decaying, suggesting that these calls are concentrated on their same gender friends. However, as men age, they get married and change their interaction preference to their opposite gender partners (see the panels for the age groups of 31-35 and 41-45 years). At the same time the distribution of call duration becomes flatter making the average call duration longer, a trend also evident among the older age groups (51-55, 61-65, and 71-75 years).  On the other hand, the ranking for the female to female (FF) calls tend to be rather low for all the age groups up to a call duration of 100 secs. The distribution of call duration is initially quite flat and small in value, but it starts increasing at about the age of women bear their first child, peaking at around 1000 secs. This suggests frequent interactions between the daughter and her mother, and seems to indicate that the grandmothering effect has set in. As for opposite gender pairs, we find that below the age of 35 years, the female to male (FM) and male to female (MF) interactions show quite high values for medium to high call duration. This can be interpreted as an indication of strong bonding between spouses. But with age, the FM-interactions start decreasing while MF-interactions increase, thus showing inverse relationship from the age of 40 years onward for medium to high call duration. This observation suggests that as women age they shift their attention from their spouses to their children.

\begin{figure}[t]
\includegraphics[width=14cm]{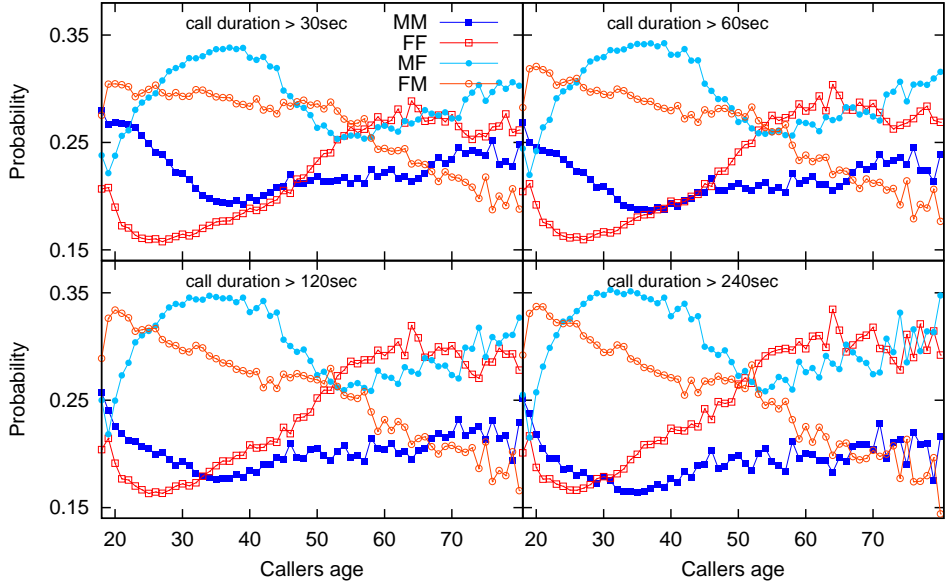}
 \caption{The relative probabilities for the four possible types of interaction for caller - callee pairs as a function of the caller age for four different values of threshold excluding calls of duration less than 30 sec, 60 sec, 120 sec and 240 sec, depicted as four panels.} 
\label{fig3}
\end{figure}

In Fig. \ref{fig3},  we depict the relative probabilities for the four caller-callee interaction categories as a function of the caller's age, for call durations greater than 30 sec, 60 sec, 120 sec, and 240 sec, respectively. It can be seen that the relative probabilities are strongly dependent on the threshold of the minimum call duration. If we concentrate only on the behavior observed for threshold values of 120 sec and 240 sec (see the two bottom panels), our observations are as follows: For individuals older than 30 years, MM interactions become less frequent, which can probably be attributed to men getting married and thus giving priority to their opposite gender spouses over the same gender friends. This picture is also supported by observing the age-wise variation of the MF-interactions, where we see that up to the age of 45 years men call their spouses more than they call others. However, MF-interactions also show a minimum around the age of 50 years after which they start increasing again from the age of 55 years on. This may be attributed to men's more frequent interactions with women one generation younger. We suggest that these are probably their daughters rather than younger female lovers, as lovers will typically be 10-15 years younger than the male, rather than 25-30 years younger as is the case here. On the other hand, the FF-interaction curve starts from a low value at about 27 years of age, after which it shows a steadily increasing trend. This observation indicates again that before marriage, females call less frequently to other females. After the age of 27, the FF-interaction curve grows rapidly up to the age of about 65 years. This behavior lends support once more to the grandmothering effect. Finally, the curve for FM-interactions indicates that after the age of 35 years, women pay less and less attention to their spouses. A similar observation also presents itself when we consider only top-ranked calls (ranked by their call duration) as shown in the Appendix, see Fig. \ref{fig6}.

\begin{figure}[t]
\includegraphics[width=14cm]{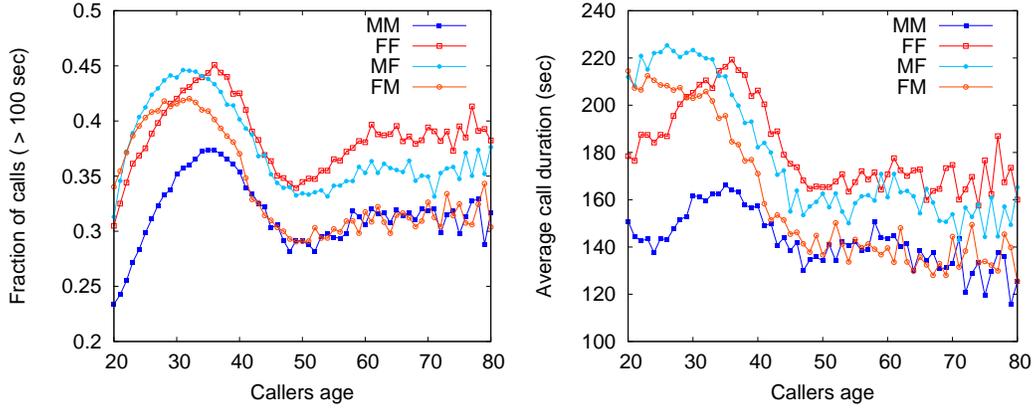}
 \caption {(Left panel) The fraction of calls of duration greater than 100 seconds as a function of the caller's age for interactions between four types of caller-callee pairs.  
 (Right panel) The average call duration as a function of the caller's age for the four types of interaction between caller and callee.} 
\label{fig4}
\end{figure}

In Fig.  \ref{fig4} (Left panel), the fraction of calls of duration greater than 100 seconds is shown as a function of the caller's age. Here, the fraction of longer calls for the four different pairs of interactions all peak around callers aged 30 years, after which the interactions decrease till about 50 years of age, followed by an increase till about 60 years of age, at which point the interactions seem to plateau.  It should be noted that, for the FF curve, the increase from 50 to 60 years of age can again be taken as clear evidence of grand-mothering. In Fig. \ref{fig4} (Right panel), we measure the average call duration as a function of caller's age for the four different types of interactions of the same or different gender pairs. From the MM curve, it is evident that the average call duration for male-to-male calls is low throughout their lifespan. The FM and MF curves show that at younger ages (i.e. before marriage) both male-to-female and female-to-male participate in long phone calls. But after typical marrying age for this population (27 years, as indicated in the national statistics), call duration drops significantly. The FF curve shows that initially the fraction increases with age (up to the age of 40 years), then rapidly falls. It is nevertheless clear that after the age of around 35 years, the call duration for female-to-female calls is the highest among the four possible types of interaction, which again can be interpreted as a signature of the grandmothering effect.

\begin{figure}[t]
\includegraphics[width=14cm]{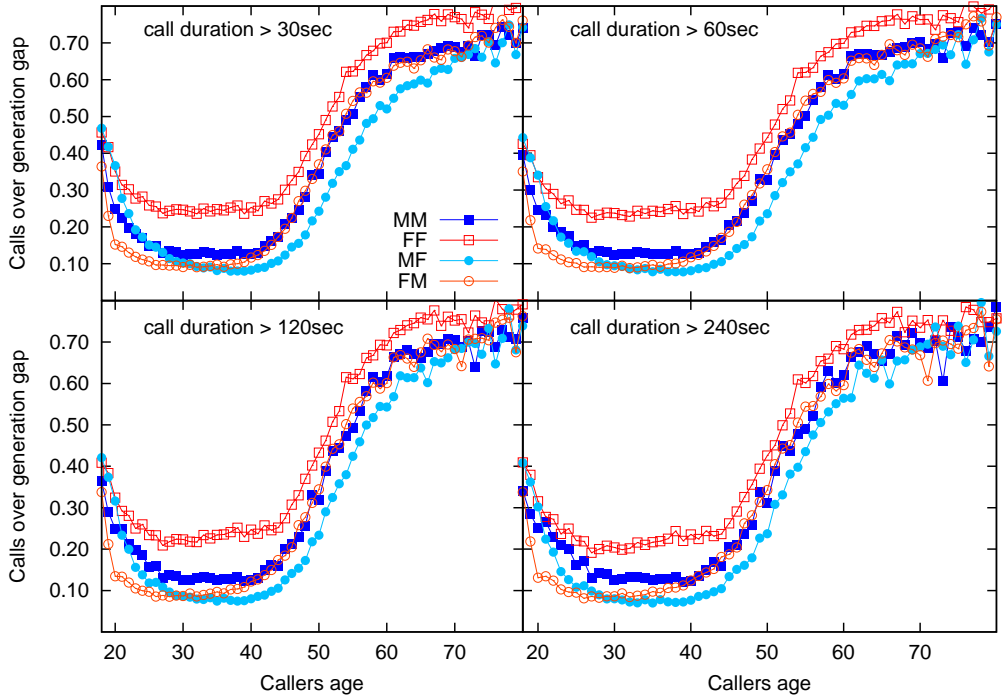}
 \caption {Fraction of calls going from the callers to previous or to next generation callees as a function of caller's age for the same and different gender pairs. The age difference between the caller and callee of about 20 years up or down is considered as an indication of a generation gap.} 
\label{fig5}
\end{figure}

In Fig. \ref{fig5}, we show the fraction of calls going from an ego to an alter who is either one generation older or one generation younger. Here we observe that FF-interactions always have the highest value for any age, which can be taken as evidence of a large amount of communication between mothers and their daughters. Before the age of 27 years (the average age of marrying in this population), measurement of MF interactions indicates that sons are also strongly attached to their mothers. After the age of 40 years, the MF and FM interactions are very close to each other, suggesting that sons get the same amount of attention from both parents. On the other hand, the tie strength between fathers and their sons are reflected in the curve for MM interactions, which show a similar trend as the other interaction types. Notice that female callers are, relatively speaking, closer to the other generation than the same age male callers. In addition, the calling patterns of older people suggest that sons and daughters get different amounts of attention from their parents. In other words, from the mothers' point of view, daughters get more attention than the sons, while sons get more attention from their fathers than daughters do. We find that, at younger ages, the fraction of calls going from one generation to another is around 10\% to 30\% of the total number of calls. On the other hand, when the age of the callers reaches 60 years, they are found to mostly communicate with their children (ranging from 50\% to 70\%), which supports the claim that \emph{affection flows downward}. A similar pattern emerges from an analysis of just the top ranked calls, as elaborated in the Appendix, see Fig. \ref{fig7}.

\section{Summary and Conclusion}
In this study, we have measured the relative interaction probabilities for the four possible caller-callee pairs of the same and opposite gender. We have observed that in general the interaction probabilities are strongly dependent on the age and gender of the caller in relation to the age and gender of the callee. Also, we observed the communication over the generation gap as depicted in Fig.\ref{fig1} showing the lobed structure and in the Appendix in Fig. \ref{fig8}, where we depict the distribution of calls made by the callers of certain age to callees as a function of callees' age showing it to be bimodal \cite{palchykov2012sex,bhattacharya2015sex,david2015communication}.  

Our findings from the study of the distributions of call duration for different age groups of the caller (Fig. \ref{fig2}) shows that the MF interactions tend to increase with call duration up to age 50 years, suggesting that men have a strong emotional connection with their opposite gender spouse of about the same age. In contrast, the FM interactions indicate that women are not as active after the age of 35 years, and have a decreasing trend for medium or long call duration with age. On the other hand, the MM interactions show initially greater probability for short call duration at younger ages, after which this becomes least probable for medium and higher call duration for any age of the caller. The FF interactions start with the lowest probability of all at younger ages, then shows a steadily increasing trend for medium and higher call duration with the age of the caller.

In the investigation of the relative probabilities for the four types of interaction as a function of the caller's age with calls above certain threshold value (30 sec, 60 sec, 120 sec and 240 sec)  (Fig. \ref{fig3}), we show that the FF interactions have an increasing trend with the caller's age. This is due to frequent interactions between the daughter and her mother, an indication that the grandmothering effect has set in. An opposite trend is observed for the FM interactions, i.e. the relative probability shows a decreasing trend with age. On the other hand, the MF interactions show a high probability for ages ranging from 20 to 50 years. After that, it shows a decreasing trend up to the age of 55 years, and then beyond that again shows an increasing trend. The MM interaction curve shows the weakest interaction after the age of 25 years.

Looking at the fractions of calls of duration more than 100 sec as a function of the caller's age (Fig. \ref{fig4}) revealed that for the FF interactions there is an increase from 50 to 60 years of age, which once again is taken as a clear evidence of grandmothering setting in. We have also found that on the basis of the average call duration that around 35 years of age the duration of female-to-female calls is highest among the four possible types of interaction. This can again be interpreted as a signature for the grandmothering effect. Furthermore, we showed (Fig. \ref{fig5}) that there is a fraction of total calls going from callers to callees who are either one generation older or one generation younger. Here we observed that for female callers the fraction of calls going to a different generation is, for all ages, always greater than for male callers. More precisely, the FF interactions show the highest probability, most likely reflecting strong ties between mothers and their daughters. On the other hand, at a younger age, a large fraction of calls go from males to females, suggesting that sons are strongly attached to their mothers before marriage. After the age of 40 years, the MF and FM interaction curves are very close to each other, suggesting that sons get the same amount of attention from both parents. More generally, we have found that at younger age, most of the calls (70-90\%) by egos are to alters of the same generation. On the other hand, for older people, most of their calls go to their children (i.e. alters a generation younger), which supports the claim that affection flows downward.

Finally notice the asymmetry in the calling pattern over the generation gap due to the fact that younger people tend not to call their parents (the older generation of alters) anything like as often as their parents call them. This suggests that parents work hard to keep contact with their children, but children do not work as hard to keep contact with their parents. At the same time, daughters are more likely than anyone else to call their mothers. In contrast, males are least likely to call their mothers or their daughters, but sons and daughters seem equally likely to call their fathers.

\section*{Acknowledgement}
A.G.  and  K.K.  acknowledge  support  from  project  COSDYN,  Academy  of  Finland  (Project  no. 276439).  K.B.,  D.M.  and  K.K. acknowledge support  from   H2020  EU  project  IBSEN. D.M. acknowledge to CONACYT, Mexico for supporting grant 383907. RD's research is supported by an ERC Advanced Investigator award.


\begin{appendices}
Here we do the same analysis as done in Fig. \ref{fig3} and Fig. \ref{fig5} by considering the top ranked calls where calls are ranked by call duration to show the robustness of long call duration.

In Fig. \ref{fig6}, we show relative probabilities for the four possible types of interaction as a function of the caller's age by considering all the calls, the top 50\% of calls, top 20\% of the calls and top 10\% of the calls (with calls ranked by duration). There are differences between all calls and top ranked calls. If we consider only potential calls, the main conclusions are (considering only top 20\% of the calls and top 10\% of the calls): (i) Beginning with a low value, the FF interaction curve has an increasing trend with caller age; (ii) the FM interaction curve starts with high value and then it shows a decreasing trend with age; (iii) the MF interaction curve shows a high value from age 30 to 40 years, after which it has a decreasing trend up to the age of 55 years, after which it shows an increasing trend; (iv) finally, the MM interaction curve differs from all the other curves by showing a low value for the probability for all the ages of the caller.

\renewcommand{\thefigure}{A\arabic{figure}}
\setcounter{figure}{0}
\begin{figure}[h]
\includegraphics[width=12cm]{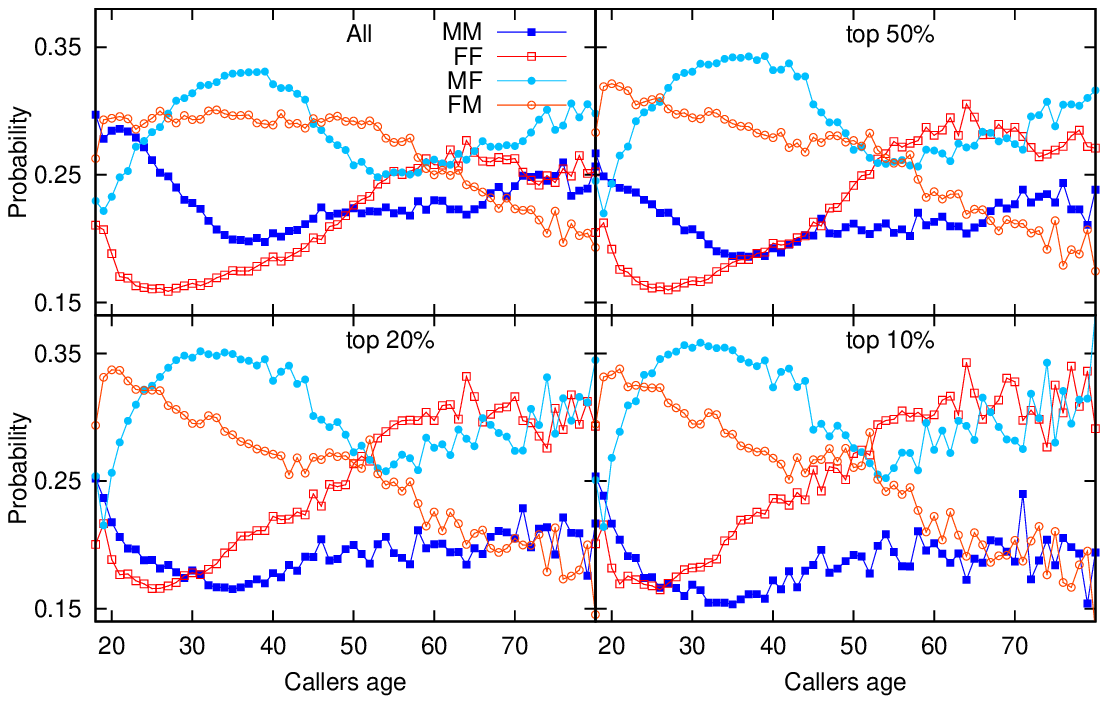}
 \caption{Relative probabilities for four possible ways of interaction as as a function the caller's age for all the calls, top 50\% of the calls, top 20\% of the calls and top 10\% of the calls, where calls are ranked by call duration.} 
\label{fig6}
\end{figure}

\begin{figure}[h]
\includegraphics[width=12cm]{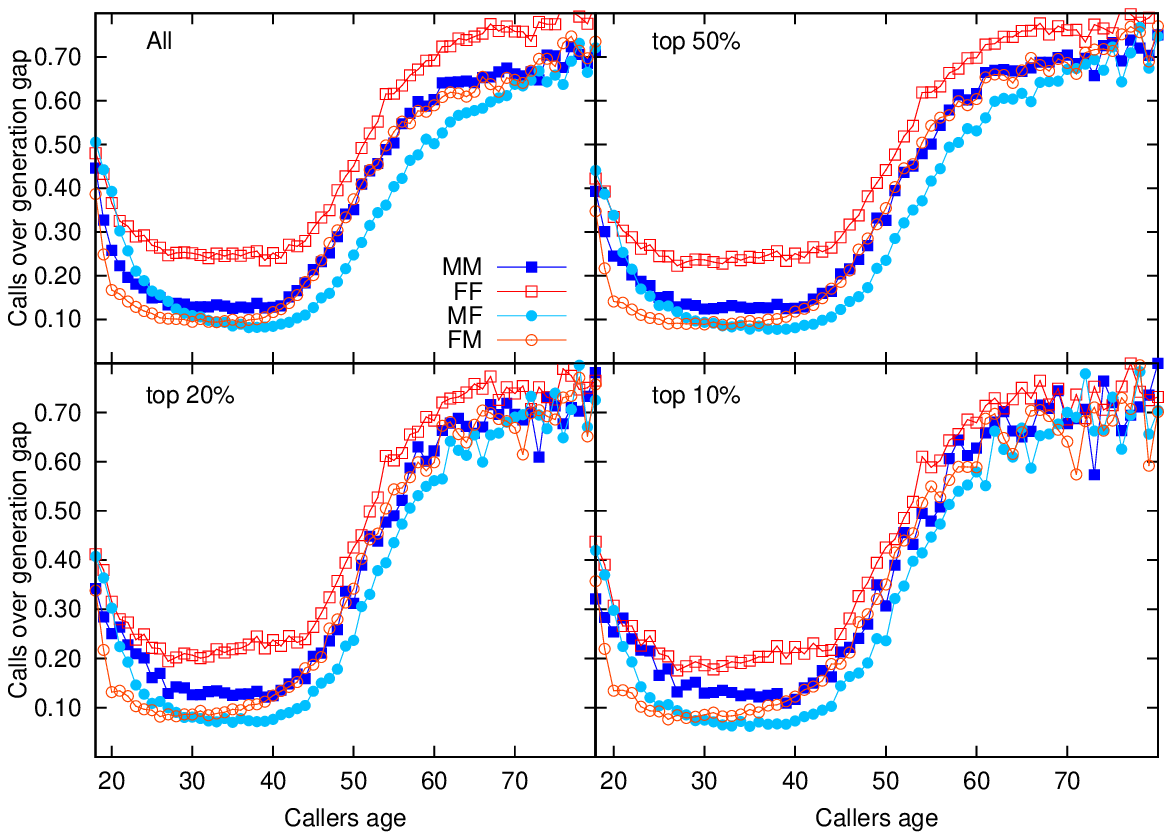}
 \caption{Fraction of the calls going from the caller to the callee of either previous or next generation by considering all the calls, top 50\% of the calls, top 20\% of the calls and top 10\% of the calls. Here the  calls are ranked by call duration.} 
\label{fig7}
\end{figure}

In Fig. \ref{fig7}, we depict the fraction of the calls going from the caller to the callee of either the previous or next generation. The main conclusions are as follows: (i) the FF-interaction curve shows the highest value for any age of the caller as clear indication of the mothers and daughters frequent connections; (ii) the MF-interaction curve indicates that sons are more attached to their mothers before they marry; (iii) after the age of 40 years, the MF and FM interaction curves are very close each other, indicating that sons get the same amount of attention from their parents. Also the behavior of all the four possible combinations of social interaction after the age of 40 tells us that mothers more frequently call their daughters than their sons, and fathers more frequently call their sons than their daughters.

\begin{figure}[h]
\includegraphics[width=16cm]{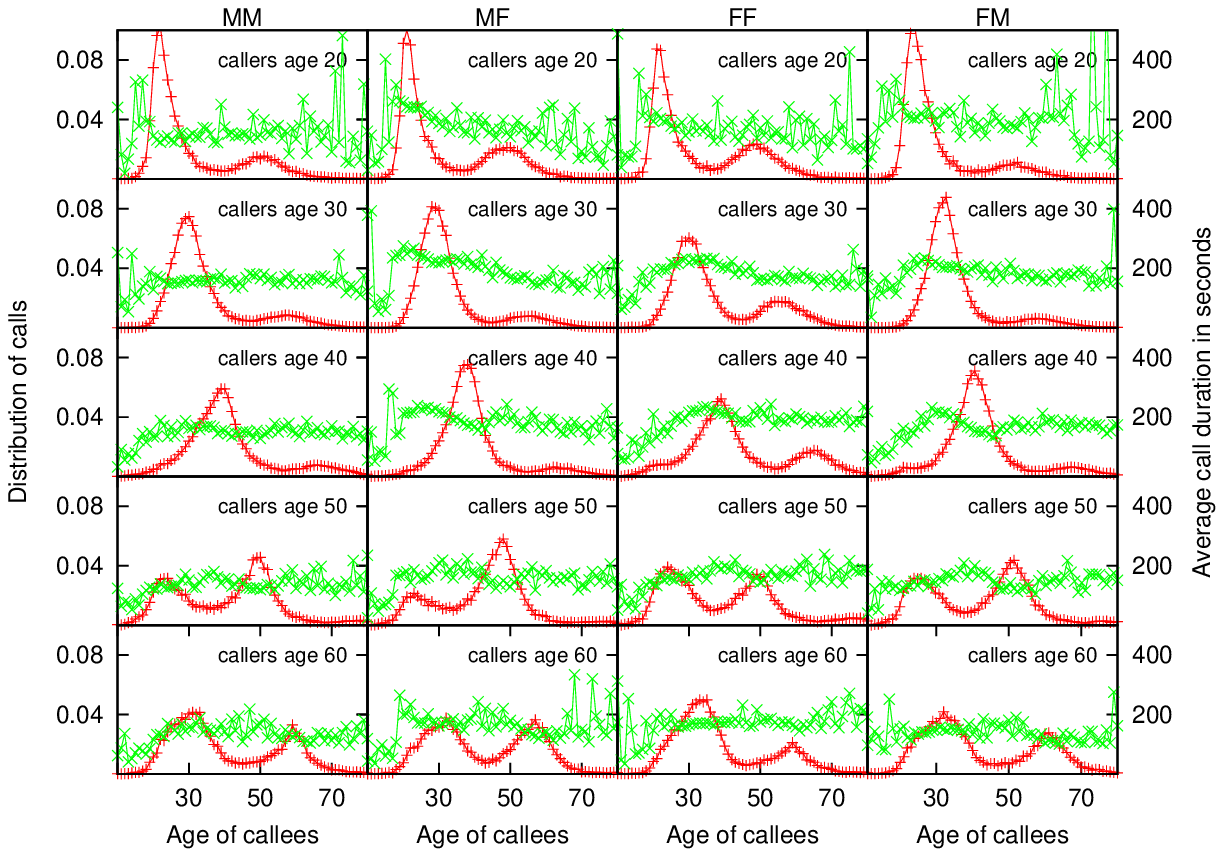}
 \caption{The distribution of calls made by the callers of different ages (i.e. 20, 30, 40, 50 and 60 years old) to callees as a function of the callees' age (red lines; scale on the left) and correspondingly the callers average call duration as a function of callees' age (green lines; scale on the right).}
\label{fig8}
\end{figure}

In Fig. \ref{fig8}, we show the distribution of calls made by the callers of certain age to callees as a function of callees age (red lines) and the corresponding average call durations (green lines). The distributions of calls turn out to be bimodal, with a maximum at around caller's own age and another maximum at an age difference of one generation \cite{palchykov2012sex,bhattacharya2015sex,david2015communication}.

\end{appendices}

\end{document}